\documentclass[a4paper,12pt]{article}
\pdfoutput=1
\usepackage{graphicx, rotating,amssymb}

\long\def\symbolfootnote[#1]#2{\begingroup\def\thefootnote{\fnsymbol{footnote}}\footnote[#1]{#2}\endgroup}
\ifx\pdfoutput\undefined
\usepackage[dvips,bookmarks]{hyperref}	
\else
\usepackage{hyperref}	
\fi
\hypersetup{colorlinks,bookmarksopen,bookmarksnumbered,citecolor=rossoc,
linkcolor=verdes,pdfstartview=FitH,urlcolor=rosso}

\def\hhref#1{\href{http://arxiv.org/abs/#1}{#1}} 
\def\mhref#1{\href{mailto:#1}{#1}}		

\usepackage{multicol}
\usepackage{color}
\definecolor{rosso}{cmyk}{0,1,1,0.4}
\definecolor{rossos}{cmyk}{0,1,1,0.55}
\definecolor{rossoc}{cmyk}{0,1,1,0.2}
\definecolor{blu}{cmyk}{1,1,0,0.3}
\definecolor{blus}{cmyk}{1,1,0,0.6}
\definecolor{bluc}{cmyk}{1,1,0,0.1}
\definecolor{verde}{cmyk}{0.92,0,0.59,0.25}
\definecolor{verdec}{cmyk}{0.92,0,0.59,0.15}
\definecolor{verdes}{cmyk}{0.92,0,0.59,0.4}

\newcommand{\beq}{\begin{equation}}
\newcommand{\eeq}{\end{equation}}

\def\Fermi{{\sc Fermi}} 
\def\PAMELA{{\sc Pamela}}
\def\AMS{{\sc Ams-02}}

\begin{document}
\begin{flushright}
\footnotesize
SACLAY--T13/137\hfill
LAPTh--019/15
\end{flushright}
\color{black}
\vspace{0.3cm}

\begin{center}
{\LARGE\bf AMS-02 antiprotons, at last!\\ [0.2cm]
Secondary astrophysical component and\\ [0.1cm]
immediate implications for Dark Matter}

\medskip
\bigskip\color{black}\vspace{0.4cm}

{
{\large\bf Ga\"elle Giesen}$^{a} \symbolfootnote[1]{\mhref{gaelle.giesen@cea.fr}, \mhref{mathieu.boudaud@lapth.cnrs.fr}, \mhref{yoann.genolini@lapth.cnrs.fr},\\ \mhref{vivian.poulin@lapth.cnrs.fr}, \mhref{marco.cirelli@cea.fr}, \mhref{pierre.salati@lapth.cnrs.fr}, \mhref{serpico@lapth.cnrs.fr}}$,
{\large\bf Mathieu Boudaud}$^{b}$,
{\large\bf Yoann G\'enolini}$^{b}$,
{\large\bf Vivian Poulin}$^{b,c}$,\\[3mm]
{\large\bf Marco Cirelli}$^{a}$,
{\large\bf Pierre Salati}$^{b}$,
{\large\bf Pasquale D. Serpico}$^{b}$
}
\\[7mm]
{\it $^a$ \href{http://ipht.cea.fr/en/index.php}{Institut de Physique Th\'eorique}, Universit\'e Paris Saclay, CNRS, CEA,\\ F-91191 Gif-sur-Yvette, France}\\[3mm]
{\it $^b$ \href{https://lapth.cnrs.fr}{LAPTh}, Universit\'e Savoie Mont Blanc, CNRS,\\ F-74941 Annecy-le-Vieux, France}\\[3mm]
{\it $^c$ Institute for Theoretical Particle Physics and Cosmology (\href{http://www.particle-theory.rwth-aachen.de/cms/Particle_Theory/~fywi/Das_Institut/lidx/1/}{TTK}), \\ RWTH Aachen University, D-52056 Aachen, Germany.}
\end{center}

\vspace{0.5cm}

\centerline{\large\bf Abstract}
\begin{quote}
\color{black}\large
Using the updated proton and helium fluxes just released by the  \AMS\  experiment
we reevaluate the secondary astrophysical antiproton to proton ratio and its uncertainties, and
compare it with the ratio preliminarly reported by \AMS. We find no unambiguous evidence for a significant excess with respect to expectations. Yet, some preference for a flatter energy dependence of the diffusion coefficient  (with respect to the {\sc Med} benchmark often used in the literature) starts to emerge.  Also, we provide a first assessment of  the room left for exotic components such as Galactic Dark Matter annihilation or decay, deriving new stringent constraints.
\end{quote}

\newpage

\tableofcontents

\section{Introduction}
Since decades, the antiproton ($\bar{p}$) component in cosmic rays (CR's) has been recognized  as an important messenger
for energetic phenomena of astrophysical, cosmological and particle physics nature (see for instance~\cite{Steigman:1976ev,Steigman:1977ej,SG87}). In modern times, antiprotons have often been argued to be an important diagnostic tool for CR sources and propagation properties, and constitute one of the prime channels for indirect searches of Dark Matter (DM)~\cite{pbar_history, DMbook}, which so far has only been detected  gravitationally. In DM annihilation (or decay) modes, antiprotons can result either from the hadronization of the primary quarks or gauge bosons or through electroweak radiation for leptonic channels.

The Alpha Magnetic Spectrometer (\AMS) onboard the International Space Station (ISS), is the most advanced detector for such indirect DM searches via charged CR flux measurements.
The positron fraction has been published earlier~\cite{Aguilar:2013qda,Accardo:2014lma}, confirming the rise at energies above 10 GeV detected previously by \PAMELA~\cite{Adriani:2008zr,Adriani:2010ib} and \Fermi~\cite{FermiLAT:2011ab}. The sum of electrons and positrons~\cite{Aguilar:2014fea} as well as their separate fluxes~\cite{Aguilar:2014mma} have also been published, thus drawing a coherent and extremely precise picture of the lepton components of CR's.
Despite the fact that DM interpretations of the positron and, more generally, leptonic `excesses' have been attempted (for a review see~\cite{Cirelli:2012tf}), even before the advent of \AMS\ it had been recognized that explanations involving astrophysical sources were both viable and favoured (for a review see~\cite{Serpico:2011wg}),  a conclusion reinforced by updated analyses (see~\cite{DiMauro:2014iia,Boudaud:2014dta}, and references therein, for recent assessments).

\medskip

In this paper, we will instead focus on CR antiprotons.
Until now, the so-called {\it secondary} antiprotons (originating from collisions of CR primaries with the interstellar material) have been shown to
account for the bulk of the measured flux~\cite{Donato:2008jk}, thus allowing to derive constraints on the DM parameter space and to compute expected sensitivities, respectively based on updated \PAMELA\ data~\cite{Adriani:2010rc} and projected \AMS\ data (see e.g.~\cite{Cirelli:2013hv,Boudaud:2014qra,Evoli:2011id,Belanger:2012ta,Fornengo:2013xda,Cirelli:2014lwa,Bringmann:2014lpa, Hooper:2014ysa}). The \AMS\ Collaboration has now presented its preliminary measurements of the $\bar p/p$ ratio~\cite{AMS2015}, with an improved statistical precision and energy range extending to $450\,$GeV. It is therefore crucial and timely to re-examine the situation and update existing results.
In addition, \AMS\ has published the measurement of the proton ($p$) spectrum~\cite{Aguilar:2015ooa} and presented the measurement of the helium ($\alpha$) one~\cite{AMS2015}, in qualitative agreement with the previous determinations by \PAMELA~\cite{Adriani:2011cu}, but now with unprecedented precision and detail. This is important for our purposes since the $p$ and $\alpha$ spectra are crucial input ingredients in the computation of the secondary antiproton flux, which is the minimal astrophysical antiproton background for indirect DM searches, as we will remind later.
Hence, with the release of these exquisitely precise datasets, \AMS\ provides a coherent, high-statistics---albeit preliminary---picture in the hadronic component of CR's too, allowing for a scrutiny of possible exotic contributions.

\medskip

However, the reach of any search for exotic physics is limited by  the astrophysical uncertainties affecting the production and the propagation processes of cosmic antiprotons in the Galaxy and in the solar system. Indeed, while the basic processes involved in the production and propagation of CR antiprotons are rather well understood, the detailed parameters entering in such processes are far from being well determined. The $\bar p$ production, propagation and Solar modulation uncertainties can have a large impact on both the astrophysical and (in particular) the DM signal. Some sensible ranges for these parameters can and must be determined  by studying ordinary CR fluxes like the ratio of Boron to Carbon (B/C ratio), which  surely have a non-exotic origin. Indeed, in this way the traditional {\sc Min}-{\sc Med}-{\sc Max} schemes~\cite{Donato:2003xg} are determined, and plausible ranges for the force field parameter of solar modulation (the so-called Fisk potential) are identified. However, these ranges are based on {\em past} CR data and are {\em not} necessarily guaranteed to work in describing the {\em current} status. We anticipate that this is what we will find in some cases discussed below. For instance, a string of recent papers, based on synchrotron radio emission~\cite{DiBernardo:2012zu,Bringmann:2011py,Orlando:2013ysa,Fornengo:2014mna} but also on positrons~\cite{DiMauro:2014iia,Lavalle:2014kca} and somewhat also on gamma rays~\cite{Fermidiffuse}, finds that the thin halo predicted by {\sc Min} is seriously disfavored.
More generally, looking for DM on top of inadequate schemes can lead to non-robust or even wrong conclusions.
Hence, one of the most crucial issues in the field is to update the uncertainty ranges of ordinary astrophysics in view of the more recent and precise experimental results, in order to build the DM search on a more solid basis. This will be possible after a careful analysis of accurate secondary over primary data like the B/C ratio, soon to be published by \AMS\ (and possibly other experiments), and provided that theoretical uncertainties will be under better control~\cite{Genolini:2015cta}. For the time being, the search for DM signatures has to be pursued with the utmost care.

\medskip

Within this broad context, the purpose of this paper is twofold: 1) based on {\em existing} propagation models, derive the state-of-the-art astrophysical antiproton background, carefully appraising the related uncertainties; 2) on the basis of such background and fully taking into account such uncertainties, assess what can be said on the room left for a DM signal, and what can not.

\medskip

The rest of this paper is organized as follows. In Sec.~\ref{background},
we remind how the computation of the astrophysical antiproton background proceeds, we detail its uncertainties and we compare the result with the measured $\bar p/p$.
In Sec.~\ref{DM} we introduce the DM contribution to $\bar p/p$ and we derive constraints on the DM annihilation cross section or decay rate, for several annihilation/decay channels and under different DM and astrophysical configurations.
Finally in Sec.~\ref{conclusions}, we conclude with a few final comments.

\section{Re-evaluation of the astrophysical antiproton background}
\label{background}

The secondary astrophysical antiproton background~\footnote{In some models one can also have a {\em primary} astrophysical source of background antiprotons, i.e.~a significant antiproton population participating to the acceleration process, see e.g.~\cite{Blasi:2009bd}.} is produced in collisions of the CR high energy protons and helium nuclei on the interstellar medium, mainly constituted of hydrogen and helium, the contributions of heavier nuclei in both projectiles and targets being a few percent correction. The locally measured flux is the result of the diffuse production in the Galactic environment and the subsequent propagation of the antiprotons to the location of the Earth.
Hence, the main ingredients of the computation for the `(secondary) astrophysical $\bar p$ source term' are: i) the injection $p$ and $\alpha$ primary fluxes from Galactic sources, ii) the collision cross sections, iii) the propagation details. While we refer to~\cite{Boudaud:2014qra,Bringmann:2006im} and reference therein for a detailed discussion of all the aspects of the computation, here we just highlight the points of novelty.

\medskip

\begin{figure}[!t]
\begin{center}
\includegraphics[width=0.495\textwidth]{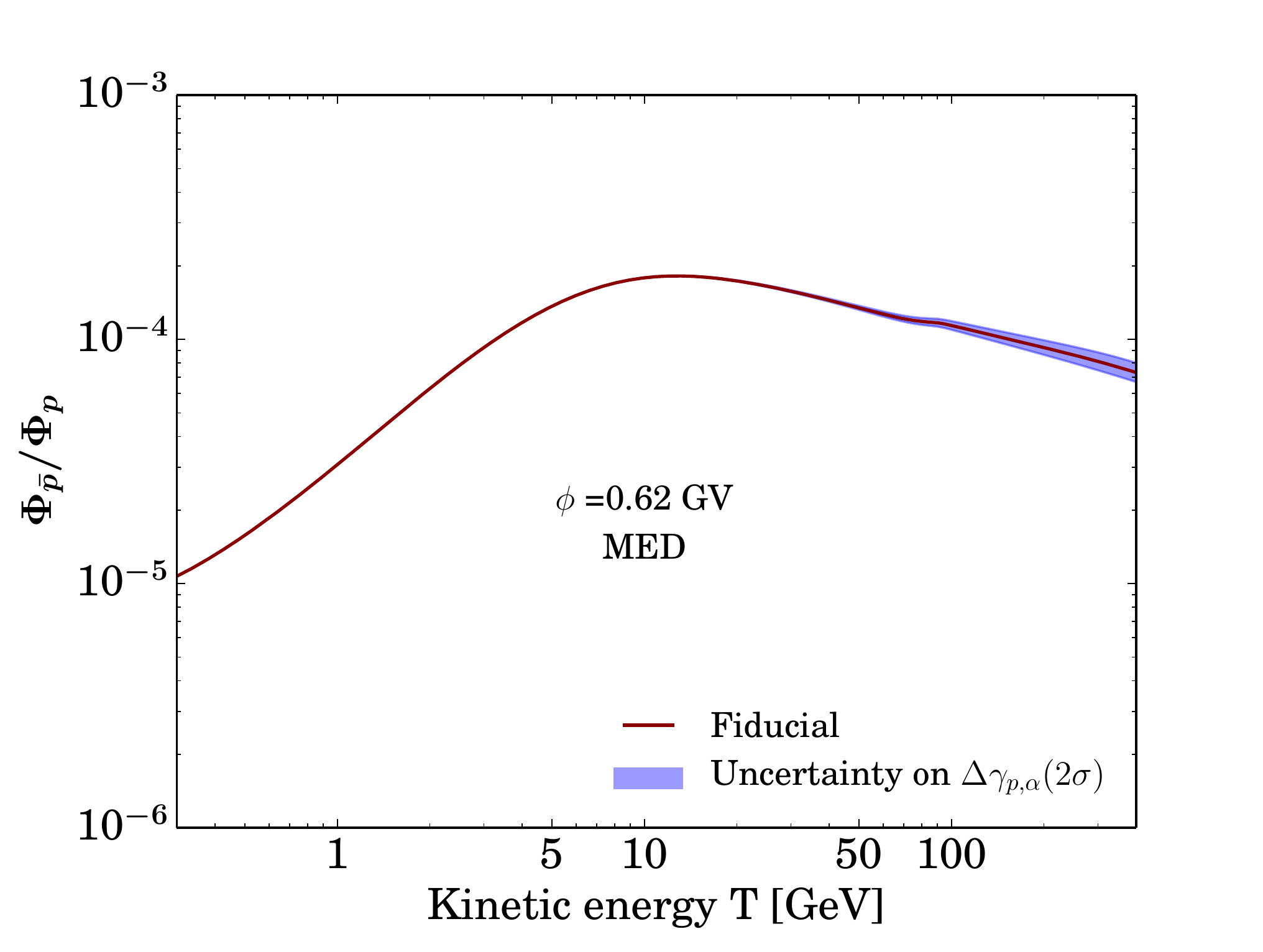}
\includegraphics[width=0.495\textwidth]{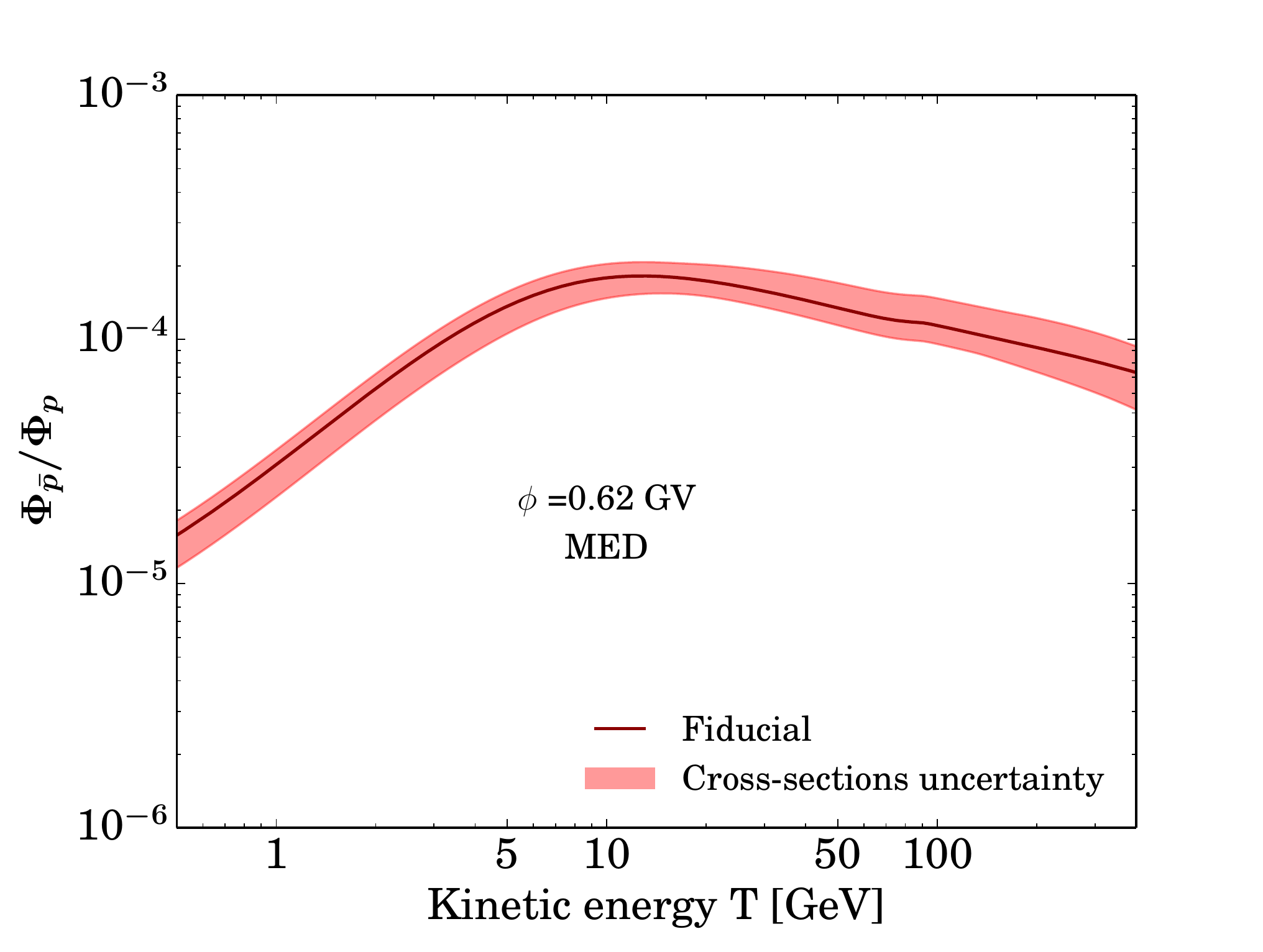}\\[0.1cm]
\includegraphics[width=0.495\textwidth]{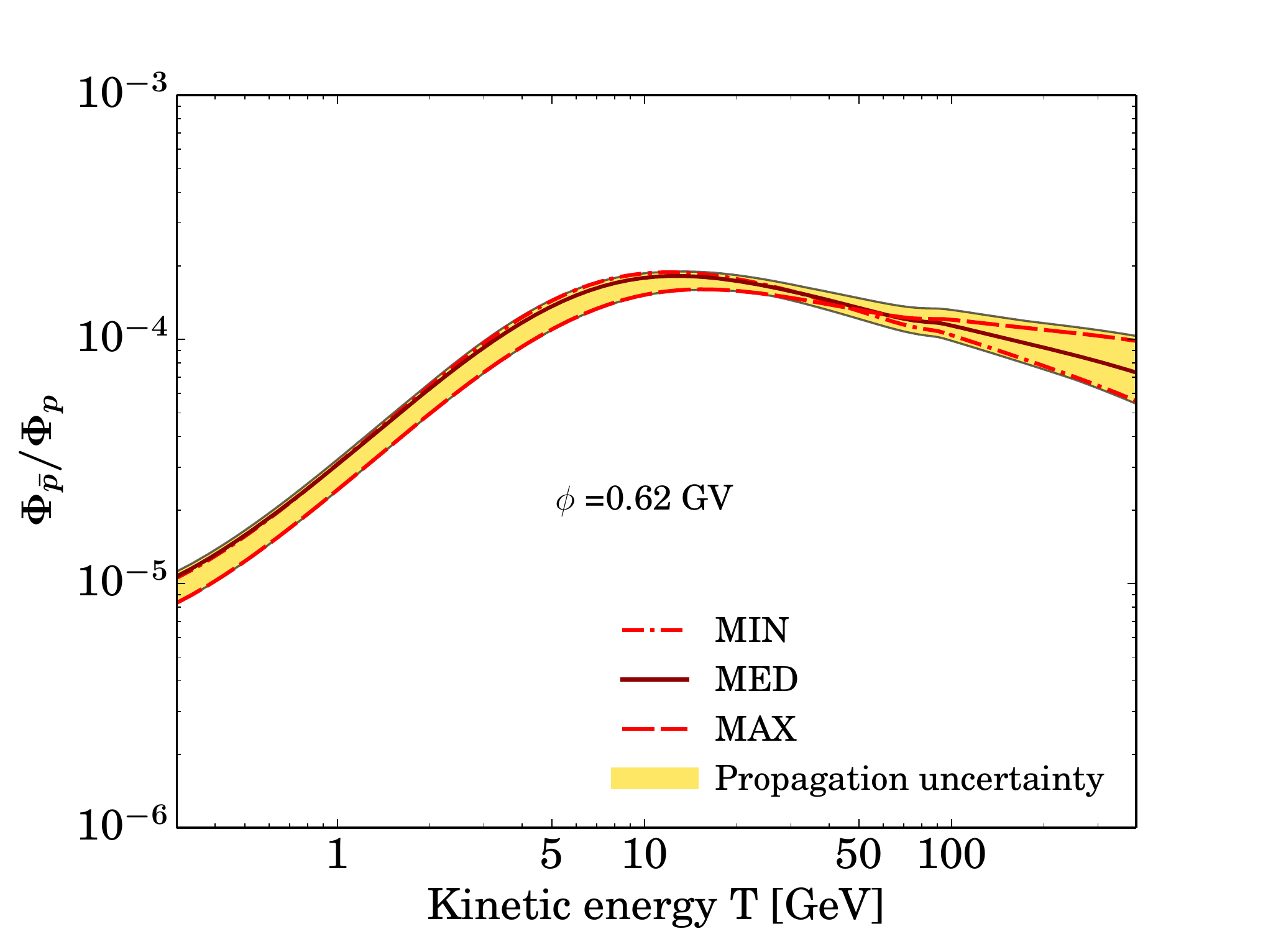}
\includegraphics[width=0.495\textwidth]{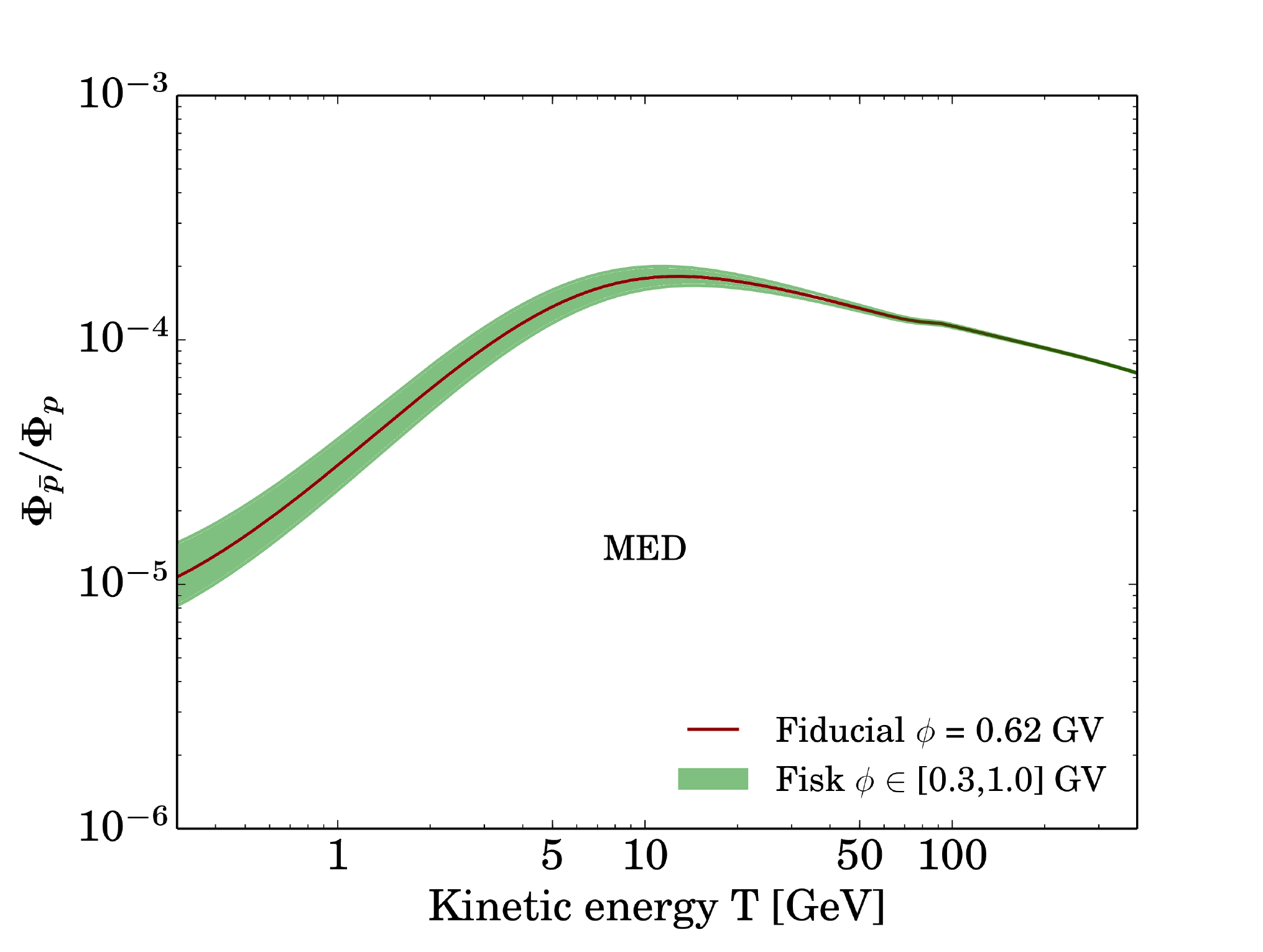}
\caption{\small \em\label{fig:background} Illustration of the individual {\bf partial uncertainties for secondary antiprotons.} The colored bands represent the uncertainties on the input $p$ and $\alpha$ fluxes (upper left panel), $\bar p$ production cross sections in the interstellar medium (upper right panel), Galactic propagation (lower left panel) and Solar modulation (lower right panel).}
\end{center}
\end{figure}

For the $p$ and $\alpha$ spectra needed in i), as mentioned above we use the data that have just been released by \AMS~{\color{rossoc}\cite{Aguilar:2015ooa,AMS2015}}. The spectra are measured up to a rigidity of 1.8 and 3 TV for $p$ and $\alpha$ nuclei, respectively, and, as already reported by the \PAMELA\ Collaboration~\cite{Adriani:2011cu}, they cannot be described by a single power law: a hardening at energies higher than $\sim$300 GV is observed for both.
At the practical level, we perform our own fits of the \AMS\ data points. The value of the Fisk potential  which gives the best $\chi^2$ for our fits is $\phi_F=0.62 \rm~GV$, the upper bound of the interval sets in \cite{Aguilar:2015ooa}. The values of the best-fit parameters are reported in appendix. The uncertainties on the slope of the {\color{rossoc}$p$} and $\alpha$ spectra  at high energies, $\Delta\gamma_{p,\alpha}$, induce an uncertainty band on the predicted astrophysical $\bar p/ p$ ratio. In fig.~\ref{fig:background}, top left panel~\footnote{Each of the panels of the figure has to assume a choice for the uncertainties presented in the other panels. E.g.~the first panel assumes definite values for the collision cross sections, a model for $\bar p$ propagation and a value for the Fisk potential. They are always chosen to be the central values, e.g {\sc Med}, the fiducial cross section and 0.62 GV for this example.}, we show the result of our computation of the ratio with such uncertainty band.
For the distribution of the sources of primary CR $p$ and $\alpha$, which can be determined from pulsar and supernova remnant surveys, we use the parameterization of~\cite{yusifov_kucuk}, slightly modified as in \cite{Bernard:2012wt}.

\medskip

For the production processes we need the cross sections $\sigma_{p{\rm H}\to\bar p {\rm X}}$, $\sigma_{p{\rm He}\to\bar p {\rm X}}$, $\sigma_{{\rm \alpha H}\to\bar p {\rm X}}$, $\sigma_{{\rm \alpha He}\to\bar p {\rm X}}$, where the first index refers to the impingent primary CR while the second one to the target interstellar material. For $\sigma_{p{\rm H}}$ we use the new parameterization recently proposed by~\cite{diMauro:2014zea}, instead of the traditional fitting relations given in~\cite{Tan_Ng_82,Tan_Ng_83}. For the cross sections of the other reactions we use the prescription of~\cite{Bringmann:2006im}, to which we refer the interested reader. We just remind that for the cross section values that we adopt the $p$H reaction dominates, providing 60\% to 65\% of the total $\bar p$ flux depending on the energy, while $p$He and $\alpha$H reactions yield 32 to 37\%, and the reaction $\alpha$He contributes less than 3\%.
Another element which has only recently been appreciated is related to the contribution of antineutron production: on the basis of isospin symmetry, one would consider the production cross section for antineutrons (e.g.~$\sigma_{p{\rm H}\to\bar n {\rm X}}$ and the others) as equal to those for antiprotons; the antineutrons then rapidly decay and provide an exact factor of 2 in the $\bar p$ flux. However, as pointed out in~\cite{diMauro:2014zea,Kappl:2014hha} and as already implemented in~\cite{Boudaud:2014qra}, it may be that this na\"ive scaling does not apply and that the antineutron cross section is larger by up to 50\% with respect to the $\bar p$ one.
Assessing uncertainties for reactions involving He is even more challenging, since {\it no data are present}, and predictions are based on semi-empirical nuclear models calibrated on data involving either protons or heavier nuclei (see~\cite{Duperray:2003bd}). For sure, uncertainties involving these reactions are at least as large in percentage as the one of the $p$H reaction, an assumption we will do in the following. More conservative assumptions would only make the error larger, and strengthen our
main conclusion on the level of agreement of the data with a purely secondary antiproton flux.
All these cumulated effects  contribute to an uncertainty band for the astrophysical $\bar p/ p$ ratio which is represented in fig.~8 of~\cite{diMauro:2014zea} and which we will adopt: it varies from about 20\% to at most 50\% (at large energies and in the most conservative conditions). In fig.~\ref{fig:background}, top right panel, we show our prediction for the $\bar p/p$ ratio with this uncertainty envelope.

\begin{figure}[!t]
\begin{center}
\includegraphics[width=0.75\textwidth]{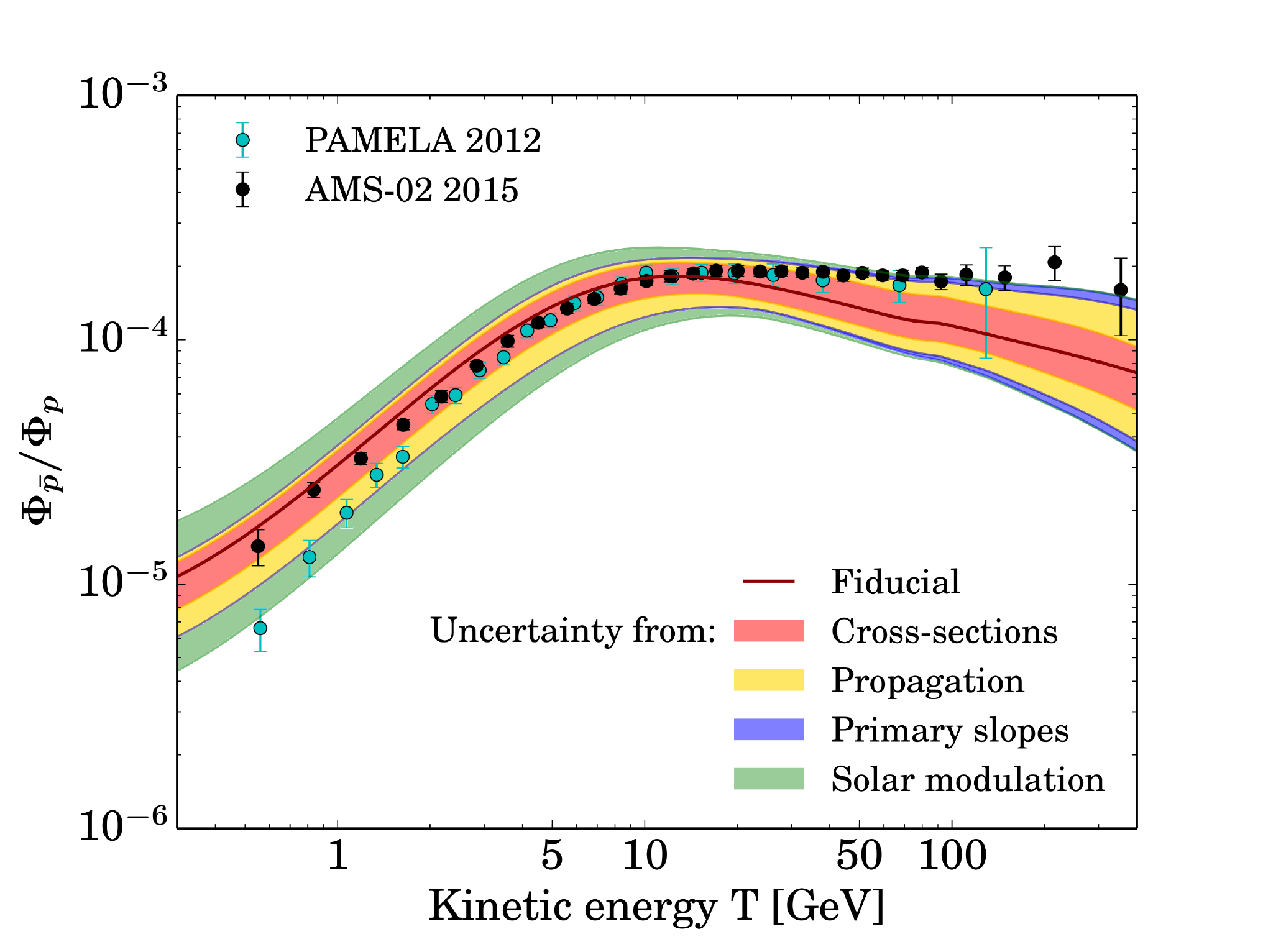}
\caption{\small \em\label{fig:background_tot}
The {\bf combined total uncertainty on the predicted secondary $\bar p/p$ ratio}, superimposed to the older \PAMELA\ data~\cite{Adriani:2012paa} and the new \AMS\ data. The curve labelled `fiducial' assumes the reference values for the different contributions to the uncertainties: best fit proton and helium fluxes, central values for the cross sections, {\sc Med} propagation and central value for the Fisk potential. We stress however that the whole uncertainty band can be spanned within the errors.}
\end{center}
\end{figure}

\medskip

Once produced, antiprotons have to propagate in the local Galactic environment before they are collected at Earth. We deal with this process in the usual way, by solving semi-analytically the full transport equation for a charged species in a 2D cylindrical `thick halo' model of the Galaxy. We do not reproduce the full treatment here (we refer again to~\cite{Boudaud:2014qra} for a self-contained description and to~\cite{Mannheim:1994sv,Strong:1998pw,ber99,Donato:2001ms,Donato:2003xg,Bringmann:2006im} for all the relevant details) but point out that we do include all the relevant processes. In particular, we take into account $\bar p$ annihilation, energy losses, `tertiary production',  and diffusive reacceleration. Besides these effects, the propagation parameters governing diffusion and convection are as usual codified in the {\sc Min}, {\sc Med} and {\sc Max} sets~\cite{Donato:2003xg}, which are by definition those that minimize or maximize a hypothetical {\it primary}, DM $\bar p$ flux at Earth. Note that these have not (yet)
been revised on the light of recent secondary data like the preliminary B/C ratio of \AMS, as discussed in the introduction, so the viability of  these predictions for the $\bar p/p$ ratio (which extends for instance to higher energies) is not trivially expected to hold.
In fig.~\ref{fig:background}, lower left panel, we show the impact of the propagation uncertainty. The curves which are labelled  {\sc Min}, {\sc Med} and {\sc Max} represent the modification which occurs by choosing these standard sets.  The shaded yellow area envelops the results obtained by sampling more widely the propagation parameter space that has been shown in \cite{Donato:2001ms} to be compatible with the B/C ratio and finding the values that minimize and maximize the {\em secondary}, rather than primary, $\bar p/p$ flux. Notice that the shaded yellow area does not coincide with the {\sc Min}-{\sc Med}-{\sc Max} envelope (see in particular between 50 and 100 GeV): this is not surprising, as it just reflects the fact that the choices of the parameters which minimize and maximize the $\bar p/p$ secondaries are slightly different from those of the primaries. However, the discrepancy is not very large.  We also notice for completeness that an additional source of uncertainty affects the energy loss processes. Among these, the most relevant ones are the energy distribution in the outcome of inelastic but non-annihilating interactions or elastic scatterings to the extent they {\it do not fully} peak in the forward direction, as commonly assumed~\cite{Donato:2001ms}. Although no detailed assessment of these uncertainties exists in the literature, they should affect
only the sub-GeV energy range, where however experimental errors are significantly larger,  and which lies outside the main domain of interest of this article.

\medskip

Finally, $\bar p$'s have to penetrate into the heliosphere, where they are subject to the phenomenon of Solar modulation (abbreviated with `SMod' when needed in the following figures). We describe this process in the usual force field approximation~\cite{Gleeson:1968zza}, parameterized by the Fisk potential $\phi_F$, expressed in GV. As already mentioned in the introduction, the value taken by $\phi_F$ is uncertain, as it depends on several complex parameters of the Solar activity and therefore ultimately on the epoch of observation. In order to be conservative, we let $\phi_F$ vary in a wide interval roughly centered around the value of the fixed Fisk potential for protons $\phi^p_F$  (analogously to what done in~\cite{Cirelli:2014lwa}, approach `B'). Namely, $\phi_F = [0.3, 1.0]\ {\rm GV} \simeq \phi^p_F \pm 50\% \, {\color{rossoc}\phi^p_F }$.
In fig.~\ref{fig:background}, bottom right panel, we show the computation of the ratio with the uncertainties related to the values of the Fisk potential in the considered interval. Notice finally that the force field approximation, even if `improved' by our allowing for different Fisk potentials for protons and antiprotons, remains indeed an ``effective'' description of a complicated phenomenon. Possible departures from it could introduce further uncertainties on the predicted $\bar p/p$, which we are not including. However it has been shown in the past~\cite{Fornengo:2013xda} that the approximation grasps quite well the main features of the process, so that we are confident that our procedure is conservative enough.

\bigskip

Fig.~\ref{fig:background_tot} constitutes our summary and best determination of the astrophysical $\bar p/p$ ratio and its combined uncertainties, compared to the new (preliminary) \AMS\ data. The crucial observation is that the astrophysical flux, with its cumulated uncertainties, can reasonably well explain the new datapoints.
Thus, our first ---and arguably most important--- conclusion is that, contrarily to the leptonic case, {\it there is no clear antiproton excess that can be identified in the first place, and thus, at this stage, no real need for primary sources}. This also means that, at least qualitatively, one expects a limited room left for exotic components, such as DM. Indeed in the following section we will proceed to compute the constraints on it.

\bigskip

\begin{figure}[!t]
	\parbox[b]{.49\linewidth}{
		\includegraphics[width=\linewidth]{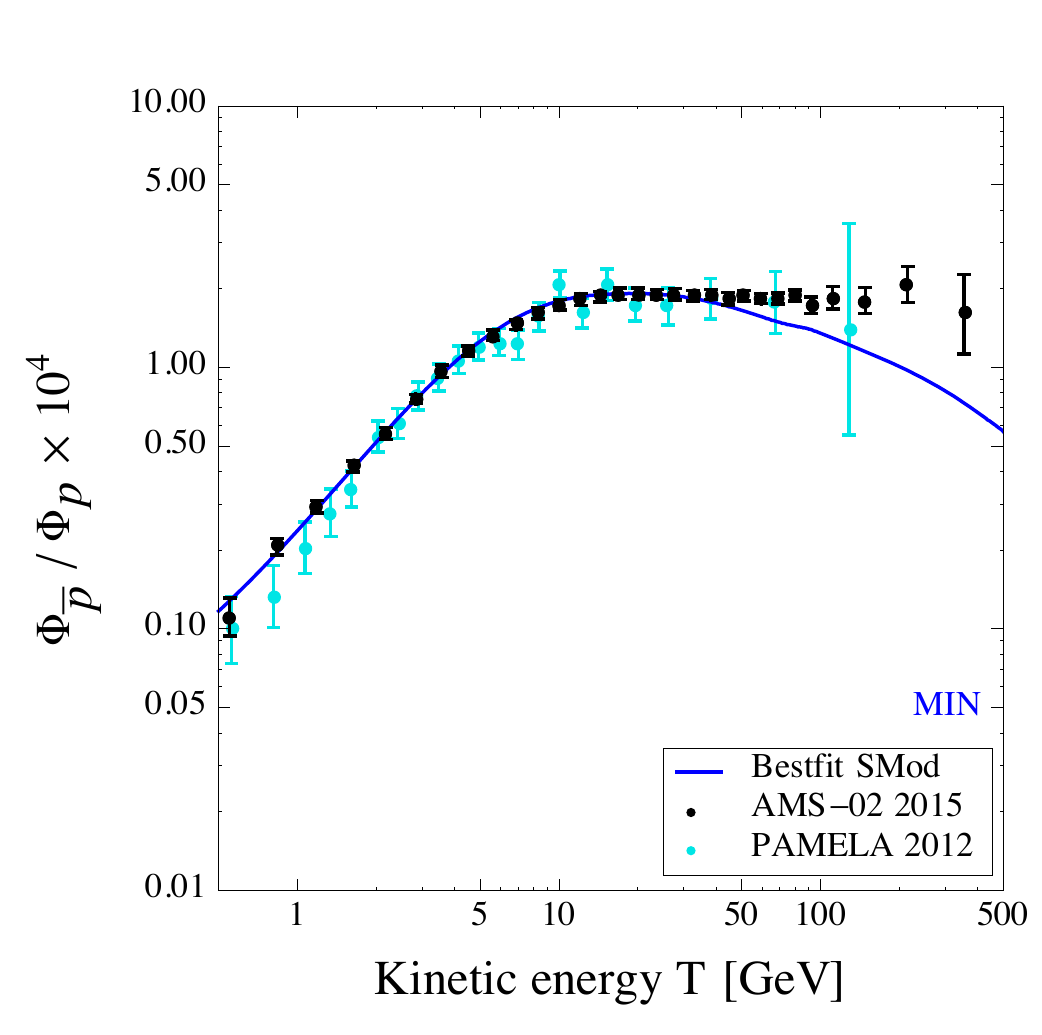}}
	\parbox[b]{.49\linewidth}{
		\includegraphics[width=\linewidth]{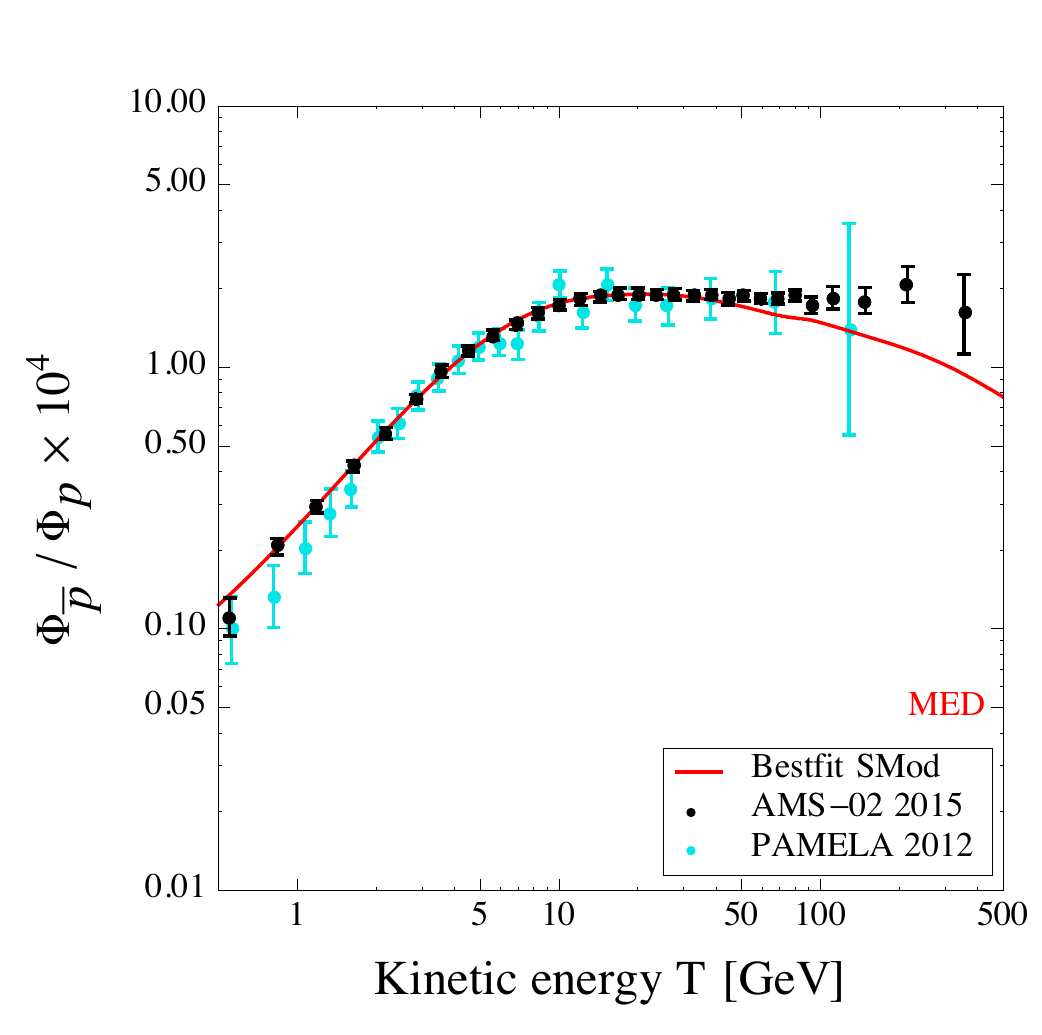}}
	\parbox[b]{.49\linewidth}{
		\includegraphics[width=\linewidth]{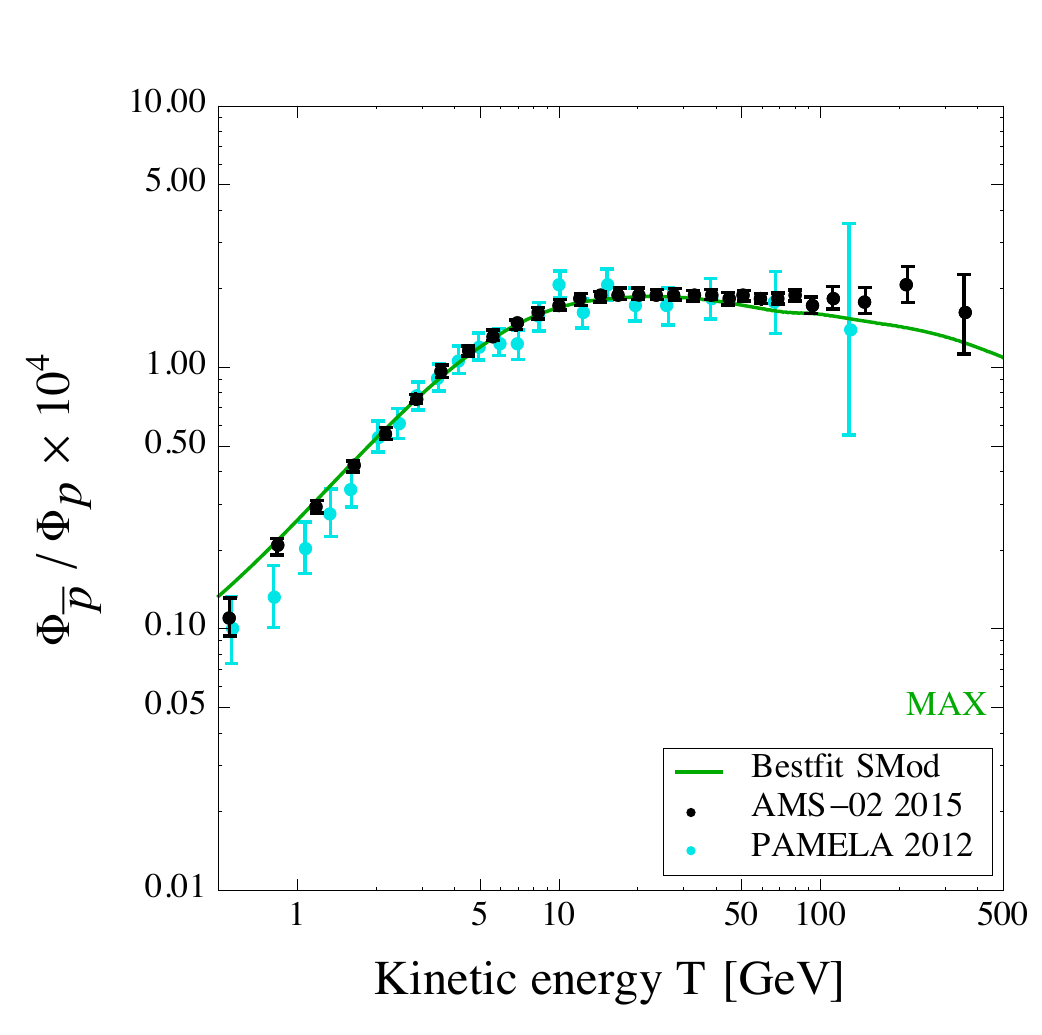}}\hfill
	\parbox[b]{.42\linewidth}{
\caption{\small \em\label{fig:background2} The {\bf best-fit secondary} antiproton fluxes originating from astrophysics, for the {\sc Min}, {\sc Med} and {\sc Max} cases, compared to the new \AMS\ data and the previous \PAMELA\ data. Each case assumes a different value for the normalization amplitude $A$ and for the Fisk potential.
{\color{white} Filling. Filling. Filling. Filling. Filling. Filling. Filling. Filling. Filling. Filling. Filling. Filling.   }}
}
\end{figure}

However, before we can do so, we have to identify specific sets of astrophysical parameters to describe the background, as discussed in the introduction. We fix in turn {\sc Min}, {\sc Med} and  {\sc Max} and we vary the Solar modulation potential in the given interval. We model the uncertainties of the production cross sections term by allowing a renormalization of the background with an energy dependence and an amplitude $A$ as dictated by the analysis presented above (namely, an uncertainty modulated as the pink band of fig.~\ref{fig:background}).
With this strategy, we look for the best fitting values of the amplitude $A$  and of the potential $\phi_F$ and we trace the corresponding $\bar p/p$ spectra.  In concrete terms, for each propagation model, we minimize the chi-square $\chi^2_0 (A, \phi_{F})$ with respect to the \AMS\ data and hence determine the best fit amplitude $A_0$ and Fisk potential $\phi_{F}^0$.
We show in fig.~\ref{fig:background2} the different cases.

Even within the limitations of the data like those we are dealing with (namely their preliminary nature, their errors only partially accounted for and the partial collection time with respect to the full lifetime of the experiment), we can see that the {\sc Min} propagation scheme predicts an astrophysical background that can {\em not} reproduce the new $\bar p/p$ data points above 30 GeV. The {\sc Med} scheme provides a barely decent fit (still good up to $\sim$ 30 GeV but rapidly degrading after) while choosing {\sc Max} the data can be well explained across the whole range of energies. We have explicitly computed the corresponding $\chi^2$ to support the above statements, with the {\sc Min}, {\sc Med}, {\sc Max} cases yielding 106, 58 and 41, respectively (for 28 degrees of freedom).  Given the preliminary nature of the data, of course they have only an indicative significance.
This is our second conclusion: {\em the preliminary $\bar p/p$ \AMS\ data seem to prefer a model, such as {\sc Max}, characterized by a relatively mild energy dependence of the diffusion coefficient at high energies}. Although it is too early to draw strong conclusions, this is an interesting observation and it goes in the same direction as the preference displayed by the preliminary B/C \AMS\ data~\cite{Genolini:2015cta}~\footnote{It is also backed by the results recently reported in~\cite{Evoli:2015vaa}---appeared after the first version of this paper in pre-print form---based on fits to \PAMELA\ B/C, $p$ and He data.}.

\medskip

It would of course be tempting to interpret the room left in the {\sc Min} and {\sc Med} cases at large energies as an exotic contribution from DM. However we insist that this would be a wrong deduction in two respects: as long as a model within the uncertainties can fit the data, failure of other models just means a better selection of the background rather than evidence for an extra component; in any case, a new assessment of the viable propagation parameter space would be needed before any conclusion is drawn.

\section{Updated constraints on Dark Matter}
\label{DM}

Primary antiprotons could originate from DM annihilations, or decays, in each point of the Galactic halo. They then propagate to the Earth subject to the same mechanisms discussed in the previous section, which are in particular described by the canonical sets of parameters {\sc Min}-{\sc Med}-{\sc Max}.
Concretely, we obtain the $\bar p$ fluxes at Earth (post-propagation) from the numerical products provided in~\cite{PPPC4DMID}, version 4. Notice that these include the subtle effects of energy losses, tertiaries and diffusive reacceleration which, as discussed at length in~\cite{Boudaud:2014qra}, are important to reach a detailed prediction.

We consider four primary annihilation (or decay) channels: DM DM $\to b\bar b, W^+W^-, \mu^+\mu^-$ and $\gamma\gamma$. These, for all practical purposes, cover very well the range of possible spectra. Indeed, annihilation (or decay) into $t\bar t$ or $hh$ (with $h$ the Higgs boson) would give spectra practically indistinguishable from those from DM DM $\to b\bar b$, and $ZZ$ from those of $W^+W^-$. The $\mu^+\mu^-$ channel represents leptonic channels, in which a small yield of antiprotons is obtained thanks to electroweak corrections (namely, the radiation from the final state leptons of a weak boson which decays hadronically). Similarly, the $\gamma\gamma$ channel produces some subdominant $\bar p$ flux via electromagnetic corrections~\footnote{For simplicity, we consider only the production of $\bar p$ from the final state. In principle, in this channel, additional hadronic production is possible from the states mediating the process of DM annihilation into photons.}.

We also consider two representative DM Galactic profiles: Einasto and Burkert, with the precise functional forms and definitions of the parameters as in~\cite{PPPC4DMID}. The former possesses a peaked distribution towards the Galactic center and hence typically results in a more abundant yield of antiprotons with respect to the latter, which features a core in the inner few kpc.

\begin{figure}[!p]
\begin{center}
\includegraphics[width=0.496\textwidth]{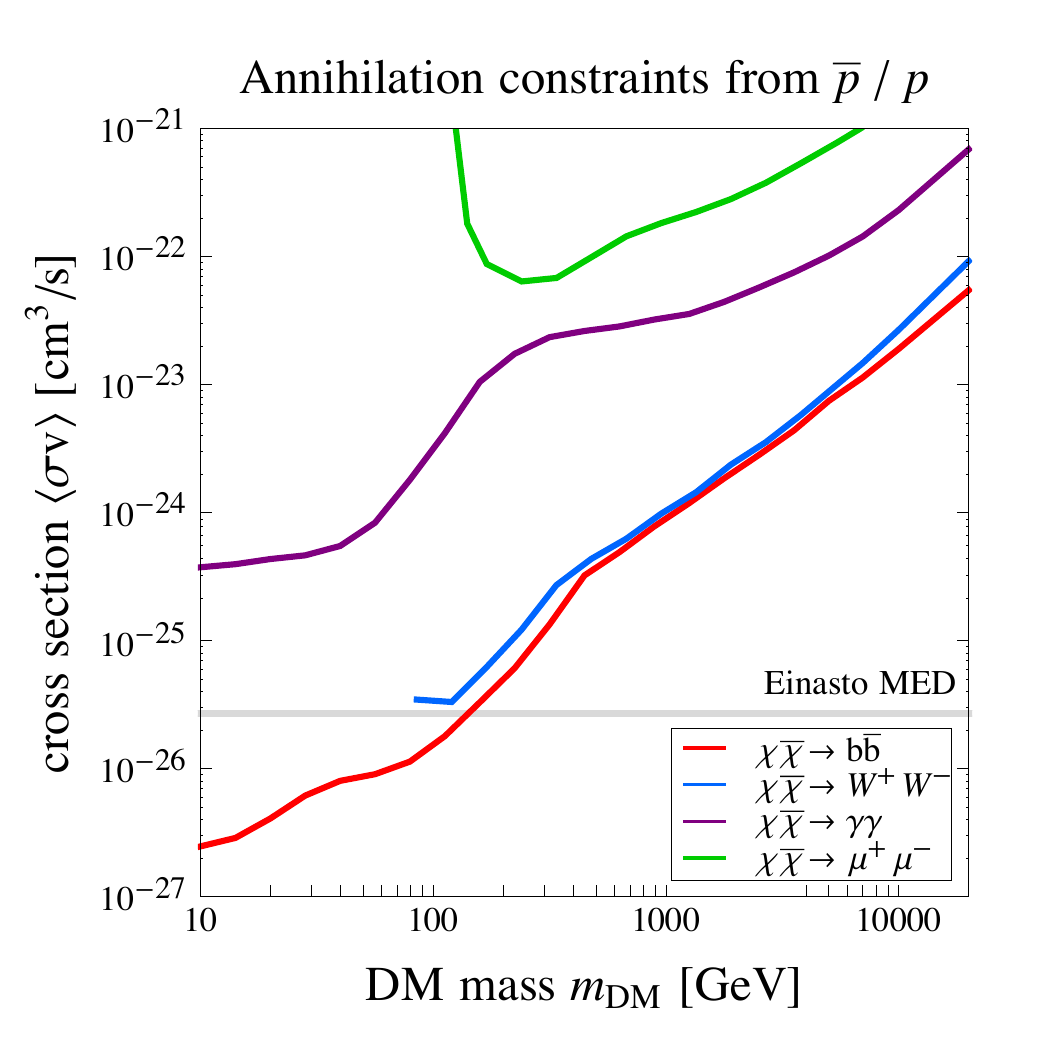}
\includegraphics[width=0.496\textwidth]{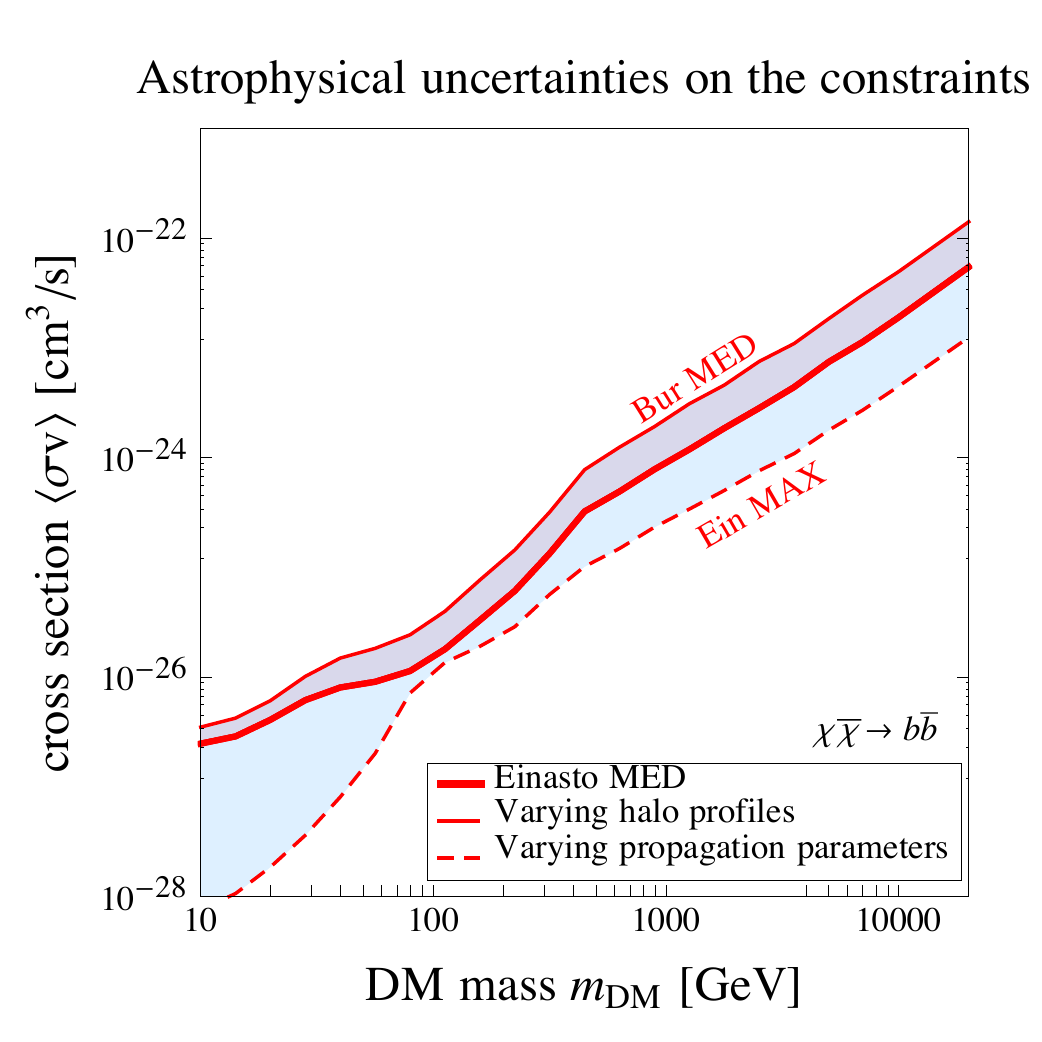}
\caption{\small \em\label{fig:constraints_ann} {\bf Annihilating DM: current constraints.} {\em Left Panel:} current constraints from the antiproton to proton ratio measurements by \AMS, for different annihilation channels. The areas above the curves are excluded. {\em Right Panel:} illustration of the impact of DM-related astrophysical uncertainties: the constraint for the $b \bar b$ channel spans the shaded band when varying the propagation parameters (dashed lines) or the halo profiles (solid lines). Notice that in the {\sc Min} case the analysis is not sensible, hence not shown here (see text for details).}
\vspace{1cm}
\includegraphics[width=0.496\textwidth]{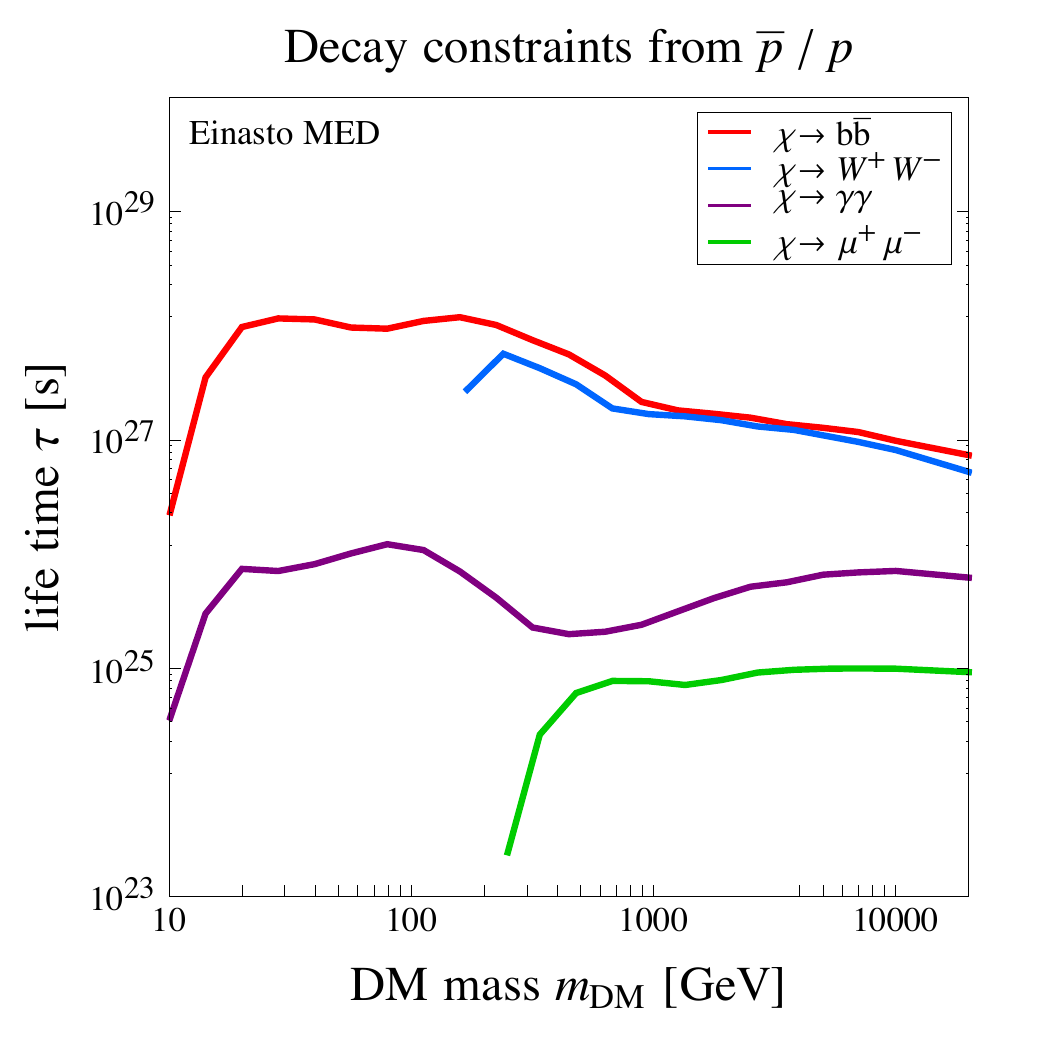}
\includegraphics[width=0.496\textwidth]{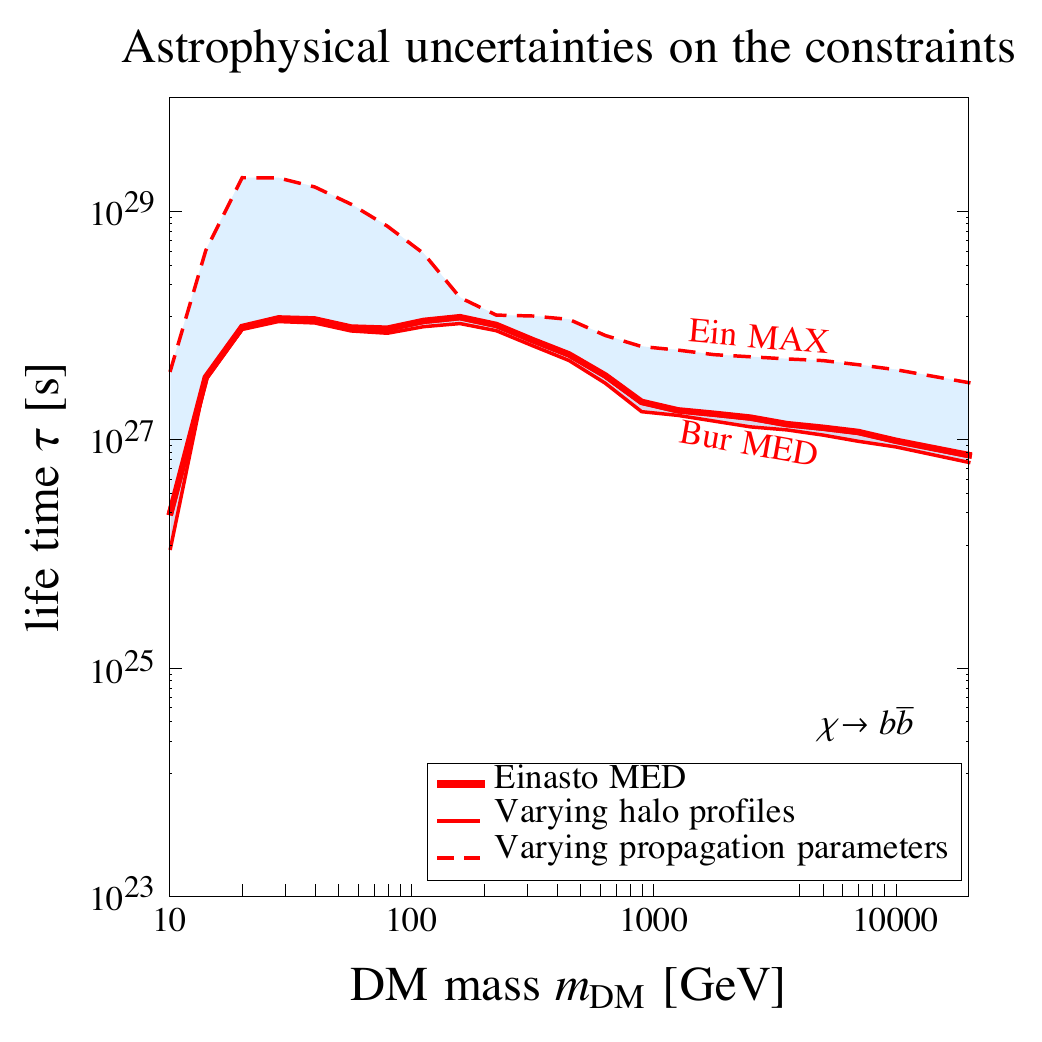}
\caption{\small \em\label{fig:constraints_dec} {\bf Decaying DM: current constraints.} {\em Left Panel:} current constraints from the antiproton to proton ratio  measurements by \AMS, for different decay channels. The areas below the curves are excluded. {\em Right Panel:} illustration of the impact of DM-related astrophysical uncertainties: the constraint for the $b \bar b$ channel spans the shaded band when varying the propagation parameters (dashed lines) or the halo profiles (solid lines). Notice that in the {\sc Min} case the analysis is not sensible, hence not shown here (see text for details).}
\end{center}
\end{figure}

\medskip

We remind that, in section \ref{background}, we have obtained the re-evaluated astrophysical background fluxes and their uncertainties. In particular, we have computed the fluxes for the {\sc Min, Med} and {\sc Max} cases, displayed in fig.~\ref{fig:background2}.
Armed with those and with the fluxes from DM as just presented, we can now compute the constraints in the usual planes `mass $m_{\rm DM}$ vs. thermally averaged annihilation cross section $\langle \sigma v \rangle$' or `mass $m_{\rm DM}$ vs. decay rate $\Gamma$'. We refer to~\cite{Boudaud:2014qra} for a detailed discussion of the practical procedure, of which we just repeat here the main lines.
Having fixed a propagation model, for a given DM mass $m_{\rm DM}$ and annihilation cross-section $\langle \sigma v \rangle$ (or decay rate $\Gamma$) we add the DM signal to the secondary background. The total flux is then
\begin{equation}
\Phi_{\rm tot} (m_{\rm DM}, \langle \sigma v \rangle, A, \phi_F)=\Phi_{\rm bkg}(A, \phi_F)+\Phi_{\rm DM}(m_{\rm DM}, \langle \sigma v \rangle, \phi_F)
\end{equation}
and we again find the best fit amplitude and Fisk potential. Finally, we solve the following equation in $\langle \sigma v \rangle$
\begin{equation}
\chi^2_{\rm DM}(m_{\rm DM},\langle \sigma v \rangle, A, \phi_F)-\chi^2_0=4,
\end{equation}
(where $\chi_0$ is the minimum chi-squared of the background-only case as computed in the previous section) in order to obtain the exclusion contour. We reproduce this for each mass point.

On the basis of the discussion in the previous section, it makes sense to derive constraints only within the propagation schemes that provide a decent explanation of the background. {\sc Max} is the favored scheme. {\sc Med} provides overall a worse but still reasonable fit to the data, so that we will employ it. In addition, (see fig.~\ref{fig:background2}, middle panel) at small energies ($T \lesssim 30$ GeV) its fit is good, thus meaningful constraints on relatively light  DM ($m_{\rm DM} \lesssim 300$ GeV) can be derived. We discard instead the {\sc Min} case.

\medskip

The results that we obtain with this strategy are presented in fig.~\ref{fig:constraints_ann} for the DM annihilation case and in fig.~\ref{fig:constraints_dec} for the DM decay case. In the left panels we fix a benchmark DM profile (Einasto) and the {\sc Med} propagation model, and show the constraints for the different particle physics channels introduced above. We see for example that the thermal annihilation cross section $\langle \sigma v \rangle = 3 \cdot 10^{-26}\, {\rm cm}^3/{\rm s}$ is now touched by the exclusion line for $m_{\rm DM} \sim 150$ GeV for the $\bar b b$ channel.
In the right panels we explore the impact of changing the propagation parameters or the DM distribution. As already highlighted several times in the literature, the effect is sizable and can reach a factor of up to an order of magnitude.
For instance, the previously quoted limit for the mass of a thermal relic can vary between 90 and 250 GeV for the range of models explored here.
Of course, as {\sc Max} maximizes by definition the DM $\bar p$ yield, its constraints are much stronger than those of the  {\sc Med} case. Turning the argument around, if the preference for {\sc Max}-like propagation schemes hinted at by preliminary \AMS\ data is confirmed, \AMS\ itself has the unprecedented possibility to exclude $m_{\rm DM} \lesssim$ 250 GeV for thermal annihilation cross section in the $\bar b b$ channel.

\section{Final remarks}
\label{conclusions}

In the light of the new $p$ flux published by \AMS\  and  the preliminary  \AMS\  results presented on the $\alpha$ flux as well as the $\bar p/p$ ratio, and using the new results of the $\bar p$ production cross sections, we have re-evaluated the secondary astrophysical predictions for the $\bar p/p$ ratio. We have accounted for the different sources of uncertainties: namely on the injection fluxes, on the production cross sections, on the propagation process and those connected to Solar modulation. Our first and main result is that
 {\em there is no unambiguous antiproton excess that can be identified in the first place, and thus, at this stage, no real need for primary sources of antiprotons}. Within errors, secondary astrophysical production alone can account for the data.  This conclusion is highly non-trivial, since we relied on updates of {\it existing} propagation schemes, which were not necessarily expected to work in the high precision and extended energy regime made accessible by \AMS.
 Adopting a more conservative treatment of the uncertainties of antiproton production cross-sections involving He as either target or projectile nuclei would clearly reinforce this conclusion.

Next, we enter in the merit of which propagation schemes do account for the data, taking into account the other uncertainties. We find that the data seem to prefer a model, such as {\sc Max}, characterized by a relatively mild energy dependence of the diffusion coefficient at high energies. If confirmed, this would go in the same direction as other indications already obtained in different channels, as discussed above.

Finally, an important application concerns updated constraints on DM: within the framework of the propagation schemes that it is sensible to use, we derive bounds that are more stringent by about one order of magnitude with respect to the previous ones~\cite{Cirelli:2013hv,Boudaud:2014qra} (based on \PAMELA\ data).

\medskip

Of course, this analysis is very preliminary and there is still room for improvements. First and foremost, the release of the final $\bar p/p$ measurement with systematic and statistical errors fully accounted for.
Yet, even a preliminary analysis allows to show that antiprotons confirm themselves as a very powerful probe for CR physics and for DM in particular. Actually, considering the puzzling excesses (with respect to the originally predicted astrophysical background) of undetermined origin in the electron and positron fluxes, considering the complicated background of most gamma-ray searches and considering the challenges of neutrino detection, $\bar p$'s might arguably still be the most promising avenue in DM indirect searches,  since improving the knowledge of the background is relatively easier than for other channels and so perhaps seeing the emergence of a clear signal is possible. In this respect, the \AMS\ experiment can play a crucial role. So far it has essentially confirmed the results of previous experiments (most notably \PAMELA), but it has done so with an impressively improved accuracy: the qualitative picture in DM indirect searches has been left largely unchanged by it, but \AMS\ has allowed improved pinning down of the parameters and tightening of the constraints. In this context, while follow-up releases of antiproton data (e.g. pure fluxes, extended energy range or enlarged statistics) will  obviously be welcome, it is urgent to address first one of the main current limitations in the field of charged CRs, namely the determination of the propagation parameters. In this respect, analyzing the upcoming reliable and accurate light nuclei measurements from \AMS\ will provide the community with a very powerful leverage for any search of exotics in CR's. At that point it will be possible to assess whether or not excesses are present in antiproton data,  for instance if the current small deficit increases in significance (although identifying their origin will remain very challenging~\cite{Pettorino:2014sua}).

\section*{Appendix: Parametrization of primary proton and helium fluxes}

To fit the \AMS\ $p$ and $\alpha$ fluxes we used the following rigidity dependent function (in particles ${\rm m}^{-2}\, {\rm s}^{-1}\, {\rm sr}^{-1}\, {\rm GV}^{-1}$):

\begin{equation}
\Phi = C \cdot (1-\beta e^{\lambda R }) \cdot \Big(\frac{R}{R+\phi_F} \Big)^2 \cdot (R+\phi_F)^{\gamma} \cdot \Big[1+ \Big(\frac{R+\phi_F}{R_{B}} \Big) ^{\frac{\Delta \gamma}{s}} \Big]^s.
\end{equation}
We proceed in two steps. First $\gamma$, $\Delta \gamma$, $R_B$, $s$ are fixed using the high energy part ($R > 45$~GV) of the spectrum. Then $C$, $\alpha$ and $\beta$ are determined over the all energy range. The value of the Fisk potential  which gives the best $\chi^2$ for our fits is $\phi_F=0.62 \rm~GV$, the upper bound of the interval sets in \cite{Aguilar:2015ooa}. The values of the best-fit parameters are reported in table \ref{table:bestfit_p_and_alpha}.

 \begin{table}[h]
     \centering
\begin{tabular}{|l|c|c|}
\hline
 & $p$ & $\alpha$   \\
\hline \hline
 C &  23566$\pm$ 30 & 4075$\pm$ 2 \\
\hline
$\lambda$ &-0.519 $\pm$ 0.007 &-0.163 $\pm$ 0.004    \\
\hline
$\beta$ & 1.21 $\pm$ 0.02 &0.41 $\pm$ 0.02    \\
\hline
$\gamma$ &-2.849   $\pm$ 0.002 &-2.795   $\pm$ 0.009  \\
\hline
 $R_{B}$&  355 $\pm$  33 & 284 $\pm$  38 \\
 \hline
$\Delta \gamma$ & 0.146 $\pm$ 0.02&0.162 $\pm$ 0.009\\
\hline
 $s$ & 0.0325$\pm$ 0.0131 & 0.078$\pm$ 0.035  \\
   \hline
$\chi^2_{ndof}$ & 29.02/(73-7) & 2.62/(54-7) \\
\hline
\end{tabular}
\caption{\small \em\label{table:bestfit_p_and_alpha} {\bf Best-fit values for $p$ and $\alpha$ fluxes.} }
\end{table}

\small
\subsubsection*{Acknowledgments}
We thank Sylvie Rosier-Lees and Antje Putze. Funding and research infrastructure acknowledgements:
\footnotesize
\begin{itemize}
\item[$\ast$] European Research Council ({\sc Erc}) under the EU Seventh Framework Programme (FP7/2007-2013)/{\sc Erc} Starting Grant (agreement n.\ 278234 --- `{\sc NewDark}' project) [work of M.C. and G.G.],
\item[$\ast$] French national research agency {\sc Anr} under contract {\sc Anr} 2010 {\sc Blanc} 041301.
\item[$\ast$] French national research agency {\sc Anr}, Project DMAstro-LHC, {\sc Anr}-12-BS05-0006.
\item[$\ast$] French {\it Investissements d'avenir}, Labex {\sc Enigmass}.
\item[$\ast$] M.C. and G.G. acknowledge the hospitality of the Institut d'Astrophysique de Paris ({\sc Iap}) where a part of this work was done.
\item[$\ast$] P.S. thanks the Institut Universitaire de France for financial support.
\end{itemize}

\bigskip

\footnotesize
\begin{multicols}{2}
  
\end{multicols}

\end{document}